\documentclass[prd,preprint,11pt,showkeys,nofootinbib]{revtex4}
\usepackage[dvips]{color}                 
\usepackage{graphicx}
\newcommand{\NPA}{\emph{Nucl.\ Phys.\ }{A}}

\newcommand{\MeV}{\mathrm{MeV}}

\newcommand{\calL}{\mathcal{L}}

\begin{document}

\title{Charged Kaon Condensation in High Density\\ Quark Matter}

\author{Paulo F.\ Bedaque}
\email{PFBedaque@lbl.gov}
\affiliation{Lawrence-Berkeley National Laboratory\\
             Berkeley, CA 94720\\
             USA} 

\begin{abstract}
We show that at asymptotically high densities the
``color-flavor-locked + neutral kaon condensate'' phase of QCD
develops a {\it charged} kaon condensate through the
Coleman-Weinberg mechanism. At  densities achievable in neutron stars
a charged kaon condensate  forms only for some (natural) values
of the  low energy constants describing the low-lying excitations of the
ground state.
\end{abstract}
\preprint{ DOE/ER/LBNL-49044}
\maketitle

%
It was realized a long time ago ~\cite{collins,barrois,bailin}
that the attraction between two quarks close to the Fermi surface
in high density strongly interacting matter leads to the formation
of Cooper pairs of quarks and the spontaneous breaking of color
symmetry. This phenomenon was more recently studied using
Nambu-Jona-Lasinio models and renormalization group methods
 ~\cite{wilczekinitial,schaferRG,cfl,hsu}, instanton models
~\cite{stonybrookinitial} and, at asymptotically high densities
where a weak coupling expansion is valid, perturbative QCD
~\cite{sonpert,pisarskipert,schaferpert}. The main lessons learned
in these studies were that i) the gap can be large, up to 100 MeV,
ii) in the case of three-flavors of quarks with the same mass the
ground state is the so-called color-flavor-locked (CFL) state ,
and chiral symmetry remains broken at arbitrarily high densities
and iii) the low-lying excitations carry the quantum numbers of
the pseudo-scalar octet familiar from the zero density case (plus
two other scalars related to the spontaneous breaking of baryon
number and axial charge) and iv) a form of electromagnetism
survives: a combination of the photon and the eight gluon is not
``Higgsed" and remains massless in the CFL phase. The equation of
state is left nearly unchanged by the pairing. A
number of electromagnetic and transport properties of quark matter
are, however, sensitively dependent on what  phase the system
finds itself and what the low energy excitation are. This
dependence provides a unique opportunity to study quark matter in
the interior of neutron stars (or rule out its existence).

 The fact that the low-lying excitation of the ground state are very
similar to the low-lying excitations of the vacuum (pions,
kaons,...) allows us to studied small perturbations around this
ground state with the techniques of chiral perturbation theory.
This brings about two advantages. At asymptotic high densities
where perturbative QCD is valid it organizes perturbative
calculations that would be very complicated otherwise. More
importantly, it provides a method to systematically expand around
the CFL phase in inverse powers of the density and/or the gap,
which is  particularly useful if  the use of perturbative QCD is not
legitimate. One of the  important perturbations around the CFL
phase is the presence of realistic quark masses. In the case of
free quarks it is easy to determine the response of the system to
these masses.  As the mass of the strange quark increases towards
its realistic value its density  decreases in such a way that its
Fermi energy equals the Fermi energy of the up quark (plus the
Fermi energy of the electrons) and the weak decay $s\rightarrow
u+e^-+\nu_e$ becomes forbidden. Charge neutrality is guaranteed by
the presence of electrons (we  assume that the neutrinos  leave
the system). The interacting case may be qualitatively different.
If the interactions are such that  quarks of different 
flavors are
paired, the change of flavor caused by flavor changing decays
 would result in
two unpaired quarks, what is not energetically favorable. As a
result, the system is rigid against small enough flavor
asymmetries.
 That is
what happens in two-flavor QCD, where up and down quarks (of two
of the three colors) are paired. An asymmetry in mass or chemical
potential between the flavors causes little change in the ground
state if they are small enough, the small change coming entirely
from the unpaired quarks of the third color
~\cite{bedaqueasymmetry}. One might think that the same effect
occurs in the three-flavor, CFL phase, since there all quarks are
paired with quarks of different flavors. The CFL phase has,
however, a another way of responding to mass asymmetries that
costs little energy but is not available in the free or two-flavor
system : it can condense mesons carrying strangeness, that are
particularly light ~\cite{bedaqueschafer}. This was demonstrated
in ~\cite{bedaqueschafer} on very general grounds and, in the case
of weak coupling, through explicit computations of the response
function to mass asymmetries. For realistic values of the quark
masses and densities  it was found that the $K^0$ is the meson
that condenses and this ground state will be referred to from now
on as the ``$CFL+K^0$" phase. For more general asymmetries,
including asymmetries on chemical potential present before weak
equilibrium is achieved and/or neutrinos leave the system, a rich
phase diagrams results, with kaonic, pionic, neutral and charged
condensates forming with different values of the parameters
~\cite{sanjaydavid}. The pattern of symmetry breaking caused by
the $K^0$ condensation (in the isospin limit) is the same one
found in the standard electroweak model $SU_I(2)\times
U_Y(1)\rightarrow U_Q(1)$ ($I=$isospin,$Y=$hypercharge and $Q=$
modified electric charge). Due to the lack of Lorentz symmetry,
only two, and not three, Goldstone bosons are generated, one
neutral and another charged
~\cite{shovkovyGoldstone,stonybrookGoldstone}. There are stable,
superconducting  topological vortices
~\cite{zhnitskyvortex,vortons} and almost stable non-topological
domains walls in the ``$CFL+K^0$" phase~\cite{sonwalls}. In the
CFL phase there is an equal number of quarks of the three flavors,
and the system achieves electrical neutrality in the absence of
any electrons, making it a perfect insulator~\cite{wilczekwrong}.
The presence of the $K^0$ condensate, being neutral, does not
change this situation. A {\it charged} kaon condensate however
would change quark matter from a perfect insulator to a
(electrical) superconductor. It is a generic feature of charged
massless scalars that the strong long wavelength fluctuations of
the gauge field lead to condensation of the scalar field
(Coleman-Weinberg mechanism (\cite{colemanweinberg})). In the
CFL+K$^0$ phase there is one {\it almost} massless charged  scalar
field. Its mass comes from isospin breaking contributions coming
from the quark mass difference and electromagnetic mass effects.
In this paper we consider the competition between the isospin
breaking mass terms and the fluctuations of the electromagnetic
field in order to determine the fate of the charged kaons and of
the possibility of a (electromagnetic) superconducting phase in
quark matter.

In the absence of quark masses, the symmetry breaking pattern
generated by diquark condensation in the CFL phase is ~\cite{cfl}
$SU_c(3) \times SU_L(3) \times SU_R(3) \times U_B(1) \times U_A(1)
\rightarrow SU_{c+L+R}(3) \times Z_2$. The electromagnetic
$U_Q(1)$ is a subgroup of  the chiral group $ SU_L(3) \times
SU_R(3) $ and the there is a surviving local
$U_{\widetilde{Q}}(1)$ ``electromagnetism" in $ SU_{c+L+R}(3) $
that is a combination of the photon and one of the gluons.
 The axial $U_A(1)$ is only an approximate symmetry of high density
 QCD due to the instanton suppression in the medium. This symmetry
 breaking pattern implies the existence of two singlet Goldstone
 bosons associated with the broken baryon number and  axial
 symmetry, and an octet of pseudoscalars. The singlets will not
 play a role in our analysis and will be dropped from now on. At
 low (excitation) energies below the gap, QCD is equivalent to the most general
 theory of an octet of pseudoscalars and photons with the same symmetries of
 QCD. This theory has been extensively analyzed
 ~\cite{casalbuoni1,casalbuoni2,seattlemass,sonstephanov,cristinalitim}
 recently. The leading terms of its lagrangian are

\begin{eqnarray}\label{lag}
  \calL&=&\frac{\epsilon}{2}\vec{E}^2 -
  \frac{1}{2}\vec{B}^2 +
   \frac{f^2}{4}{\rm Tr}[D_0\Sigma^\dag D_0\Sigma - v^2 \nabla \Sigma^\dag
   \nabla\Sigma]
   + \frac{a\Delta^2}{8 \pi}{\rm Tr}[\widetilde{{\mathcal{M}}}
   (\Sigma+\Sigma^\dag-2)]\nonumber\\
   &+& b\ \tilde{\alpha}f^2 \Delta^2{\rm Tr}[\Sigma^\dag,Q][Q,\Sigma]
   + c \ \tilde{\alpha}^2 f^4 \ ({\rm Tr}[\Sigma^\dag,Q][Q,\Sigma])^2+\ldots .
\end{eqnarray}
\noindent where $\mu$ is the baryon chemical potential, $\Delta$
is the gap, $D_0\Sigma=\partial_0 \Sigma +
\frac{i}{2\mu}[{\mathcal{M}}^2,\Sigma]-i \widetilde{e}A_0
[Q,\Sigma]$, $\vec{D}\Sigma=\vec{\nabla}_0 \Sigma -i
\widetilde{e}\vec{A} [Q,\Sigma]$, $\widetilde{e}= e g
\sqrt{3}/\sqrt{3g^2+4 e^2} $ ($g=$strong coupling constant, $e$
the electron charge), $\widetilde{\alpha}=\widetilde{e}^2/4\pi$,
${\mathcal{M}}$ and $Q$ are the quark mass and charge matrix,
 $\widetilde{{\mathcal{M}}}={\mathcal{M}}^{-1}{\rm
 det}{\mathcal{M}}$ and  $\Sigma=e^{i \pi^A \frac{\lambda^A}{\sqrt{2} f}}
 $. A few comments are in order here. The electromagnetic field in
 Eq.~(\ref{lag}) are the rotated fields that remain massless in
 the CFL phase.
 The low energy constants $f,v,a,b,c,\epsilon$ can,
 in principle, be determined from QCD. In practice, this can be
 done only in the asymptotic limit where perturbation theory is
 valid. At lower densities one can estimate their values
by looking at their variation with the cutoff of the effective
theory (see below).
 The dielectric constant $\epsilon$ was computed
 in ~\cite{cristinalitim} where it was found that
  $\epsilon=1+\frac{8}{9\pi}\frac{\widetilde{\alpha}
  \mu^2}{\Delta^2}$. The magnetic permeability was argued to
  be unchanged from the vacuum value because the diquark
  condensate carries no spin. We will assume this to be true even
  outside the perturbative QCD regime. The values of $f$, $v$
   were also determined in perturbation theory
   ~\cite{sonstephanov} to be $v=1/\sqrt{3}$ and
   $f=(21-8\ln(2))/36\pi^2$. After some controversy
   ~\cite{casalbuoni2,sonstephanov,seattlemass,sonstephanoverratum,thomasmass}
   the value of
   $a$ seems to have settled at $a=12/\pi$~\cite{sonstephanoverratum,thomasmass}.
   Finally, the value of the gap is estimated to be around $50-100$ MeV in phenomenological
   models at $\mu\simeq 500$MeV and is given in perturbation theory
   by  $\Delta= 512(2/N_F)^5/2\pi^4/g^5 e^{-\frac{3\pi^2}{\sqrt{2}g}}$
   ~\cite{sonpert,pisarskipert}, although large corrections from
   higher orders are expected
   ~\cite{rockfeller,seattleimprovement,dirk}. The coefficient of the term quartic in
   ${\mathcal{M}}$ in Eq.~(\ref{lag}) is not a free parameter because
   it is related by an approximate local symmetry of high density QCD to the
   kinetic term~\cite{bedaqueschafer}. Terms violating this
   symmetry (like the other mass term in Eq.~(\ref{lag})) are suppressed by extra
   powers of $1/\mu$. The electromagnetic coefficients $b$ and $c$
   have not yet been computed in perturbation theory but we will
   estimate them below. The terms implied by the dots in
   Eq.~(\ref{lag}) are further suppressed by powers of momenta or
   meson masses in units of the cutoff $\Lambda\simeq 2 \Delta$ or extra powers
   of $\widetilde{\alpha}$.

   For values of $\Delta$ satisfying
\begin{equation}\label{rangeDelta}
  \cot\left( \frac{\phi}{\sqrt{2}f}\right)
\frac{f^2}{\mu^2}\ \frac{\pi (m_u+m_d)m_s}{a-
  \frac{64\pi b\widetilde{\alpha} f^2}{m_s(m_d-m_u)}}
  < \Delta^2
  < \frac{f^2}{\mu^2}\frac{\pi}{a}\frac{m_s^3}{m_u}.
\end{equation}
\noindent the minimum of
   the potential is found at ~\cite{bedaqueschafer}
\begin{equation}\label{k0ansatz}
  \Sigma=\left(
  \begin{array}{ccc}
    1 & 0                              & 0 \\
    0 & \cos(\frac{\sqrt{2} \phi}{f})  &  i\sin(\frac{\sqrt{2} \phi}{f})\\
    0 & i\sin(\frac{\sqrt{2} \phi}{f}) &  \cos(\frac{\sqrt{2} \phi}{f})
  \end{array}
  \right),
\end{equation}
\noindent with
\begin{equation}\label{phiiso}
\cos(\frac{\sqrt{2} \phi}{f})=\frac{a m_u\Delta^2 \mu^2}{\pi
   f^2(m_s-m_d)^2(m_s+m_d)},
\end{equation}
\noindent describing a $K^0$ condensate (for reasonable values
of the parameters $\cot\left( \frac{\phi}{\sqrt{2}f}\right) $ is nearly one).
 The upper limit in
Eq.~(\ref{rangeDelta}) is the maximum value of $\Delta$ for $K^0$
condensation and the lower limit marks the onset of $K^+$
condensation. At very high $\mu$ the numerical value of the range in
Eq.~(\ref{phiiso}) is fairly independent of the chemical potential
$\mu$. Taking the perturbative QCD values of $f,a,v$ and
$\epsilon$,  $m_u=4$ MeV and $m_s=150$ MeV this range is
 $(2\MeV)^2< \Delta^2 < (120\MeV)^2)$
and  thus, most likely, the real world case will
correspond to the ``CFL+K$^0$" phase. Notice that in the CFL phase
the kaons are the lightest mesons and so are the first to condense
under perturbations. In the isospin limit  neutral and charged
kaons are degenerate. The reason for the neutral kaons to condense
are their slightly smaller mass due to the $m_d-m_u$ quark mass
difference and electromagnetic corrections and the fact that the
presence of a charged kaon condensate implies  the presence of
electrons to guarantee charge neutrality, which raises the energy
of this state compared to the CFL+K$^0$ phase.

\begin{figure}[t]
\centerline{\includegraphics*[bbllx=38,bblly=517,bburx=266,bbury=710,scale=0.95,clip=true]{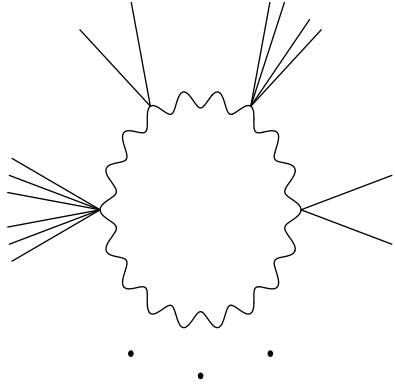}}
\caption{ \label{fig:graphs}\textit Graphs giving rise to the one loop effective
potential in the Landau gauge. Solid lines are mesons, wiggly line
are photons. }
\end{figure}

This conclusion may be changed by the inclusion of photon loops
and electromagnetic interaction terms that, by consistence, must be included together.
 Let us now compute the one photon
loop contribution to the effective potential and leave the
discussion of the conditions under which it is important for
later.

It is convenient to use a modified Landau gauge fixing procedure,
that is, we add a term $-1/2\xi (\partial_0 A^0+v^2
\nabla\overrightarrow{A})^2$  to the lagrangian, taking the limit
$\xi\rightarrow 0$. The price payed by having a complicated
propagator that breaks Lorentz invariance is compensated by the
fact that all zero external momentum one-loop diagrams involving a
meson propagator vanish, since the photon propagator satisfies, in
this gauge

\begin{equation}\label{landaugauge}
  p_\mu V^{\mu\nu}D_{\nu\lambda}(p)=0, \ \ \ \ \ \ V^{\mu\nu}={\rm
  diag}(1,v^2,v^2,v^2).
\end{equation}

 The one-loop
effective potential is then given by the sum of diagrams shown in
Fig.(\ref{fig:graphs}):

\begin{eqnarray}\label{oneloopcomputation}
  V_{1-{\rm loop}}&=&\sum_{n=1}^\infty  \frac{1}{2n}
  \left(- \frac{\widetilde{e}^2 f^2{\rm
  Tr}[\Sigma^\dag,Q][Q,\Sigma]}{2}\right)^n
  \lambda^{-\eta}\int\frac{d^Dp}{(2\pi)^D} {\rm Tr} \underbrace{D(p)V
  D(p)V\ldots}_n\nonumber\\
&=&-\frac{i}{2}\lambda^{-\eta}\int\frac{d^Dp}{(2\pi)^D} \left[ 2
\ln\left( 1- \frac{v^2 \widetilde{e}^2 f^2{\rm
  Tr}[\Sigma^\dag,Q][Q,\Sigma]/2 }{\epsilon p_0^2-p^2}
\right)\right. \nonumber\\
 & &\phantom{=-\frac{i}{2}\lambda^{-\eta}\int\frac{d^Dp}{(2\pi)^D}
             }
 \ \ + \left. \ln\left( 1- \frac{v^2 \widetilde{e}^2 f^2{\rm
  Tr}[\Sigma^\dag,Q][Q,\Sigma]/2 }{\epsilon(p_0^2-v^2 p^2)}
\right)  \right]\nonumber\\
&=&\frac{1}{64\pi^2}(\frac{2
v^4}{\sqrt{\epsilon}}+\frac{v}{\epsilon^2}) \left(\frac{
\widetilde{e}^2 f^2{\rm
  Tr}[\Sigma^\dag,Q][Q,\Sigma]}{2}\right)^2
   \lambda^{-\eta}\nonumber\\
  & &\phantom{\frac{1}{64\pi^2} }
  \left[(\frac{2}{\eta}+\gamma-\ln
  4\pi-\frac{3}{2})+\ln\left( \frac{v^2 \widetilde{e}^2 f^2{\rm
  Tr}[\Sigma^\dag,Q][Q,\Sigma]}{2} \right) \right],
\end{eqnarray}
\noindent where we work in $D=4+\eta$ dimensions and $\lambda$ is
an arbitrary renormalization scale.
The ultraviolet divergences
are absorbed in the terms proportional to $b$ and $c$ in
Eq.~(\ref{lag}), which suggests that the natural values for these
(renormalized) constants at the cutoff scale $\lambda\simeq
2\Delta$ are
\begin{eqnarray}\label{NDA}
  b(\lambda\simeq 2\Delta)&=&
  \frac{\bar{b}}{8\pi}(\frac{2
v^2}{\sqrt{\epsilon}}+\frac{1}{v\epsilon^2}),\nonumber\\
c(\lambda\simeq 2\Delta)&=&\frac{\bar{c}}{16} (\frac{2
v^4}{\sqrt{\epsilon}}+\frac{v}{\epsilon^2}),
\end{eqnarray}
\noindent where $\bar{b}$ and $\bar{c} $ are numbers of order
one\footnote{The values of $b$ and $c$ are gauge and
renormalization prescription dependent. We refer here to their
values in the Landau gauge (\ref{landaugauge}) and in the modified
minimal subtraction scheme with renormalization scale
$\lambda\simeq 2 \Delta$ }. In addition we expect on physical
grounds that $\bar{b}<0$, what guarantees a positive
electromagnetic contribution to the mass square for the mesons
(the photon loop contribution vanishes in the Landau gauge). This
estimate of the coefficient $b$ agrees with the ones in 
~\cite{dkhongelectro,tytgat} (where no attempt was made to count factors of
$4\pi$) and 
~\cite{sanjaydavid} (where no attempt was made to count factors of
$\epsilon$ or $v$). Other electromagnetic terms not renormalized
at one loop order are assumed to be suppressed. Also, the
contribution from meson loops is proportional to $(m_K/f)^4\sim
(\Delta/\mu)^4 ((m_u+m_d)m_s/f)^2 $ and is strongly suppressed.
 Finally, the
effective potential including the one photon loop correction
becomes
\begin{eqnarray}\label{Veff}
  V_{eff}&=&-\frac{a \Delta^2}{8 \pi}{\rm Tr}\widetilde{{\mathcal{M}}}
   (\Sigma+\Sigma^\dag-2)
   -\frac{f^2}{16\mu^2}{\rm
   Tr}[\Sigma^\dag,{\mathcal{M}}^2][{\mathcal{M}}^2,\Sigma]\nonumber\\
   &-&\bar{b}
\left(\frac{2 v^2}{\sqrt{\epsilon}}+\frac{1}{v\epsilon^2} 
  \right) \frac{\widetilde{\alpha} \Delta^2 f^2}{8\pi}  {\rm
  Tr}[\Sigma^\dag,Q][Q,\Sigma]\\
  &+&(\frac{2
v^4}{\sqrt{\epsilon}}+\frac{v}{\epsilon^2}) \frac{
\widetilde{\alpha}^2
  f^4}{16}
   ( {\rm
  Tr}[\Sigma^\dag,Q][Q,\Sigma] )^2
  \left[ \ln\left( \frac{v^2 \widetilde{e}^2 f^2 {\rm
  Tr}[\Sigma^\dag,Q][Q,\Sigma]}{8 \Delta^2} \right)-\bar{c}-\frac{3}{2}\nonumber  \right].
\end{eqnarray}
Whether the electromagnetic corrections computed above can modify
the position of the minimum of the potential depends on the
hierarchy assumed for the  scales $\mu$, $\Delta$, etc.

It is probably instructive to compare the present situation with
the simpler one  of scalar QED. The effective potential there has
the form (omitting numerical factors) \cite{colemanweinberg}
\begin{equation}\label{scalarQED}
  V=m^2 \phi^2 + \lambda \phi^4 +(\lambda^2+\alpha^2) \phi^4\left(
  \frac{\alpha\phi^2}{M^2}\right).
\end{equation}
In the massless case the minimum of (\ref{scalarQED}) is at
$\bar{\phi}^2=M^2/\alpha\
e^{-\frac{1}{2}+\frac{\lambda}{\alpha^2}}$. Assuming $\lambda\alt
\alpha^2$ the term proportional to $\lambda^2$ can be disregarded
and higher loop corrections are under control at
$\phi=\bar{\phi}$. The presence of a finite mass term will not
destroy this minimum if, at $\phi=\bar{\phi}$, it is smaller than
the other terms. This condition translates into $m^2
e^{-\frac{1}{2}+\frac{\lambda}{\alpha^2}}< \alpha M^2$. In our
case, we have mass terms of the order $\alpha \Delta^2$, $
(m_d-m_u)m_s \Delta^2/\mu^2$, and $M^2\sim \Delta^2$. The role of
the self-interaction is played by terms coming from the
electromagnetic mass term ($\lambda\sim \alpha \Delta^2/\mu^2$),
the quark mass term ($\lambda\sim (m_d-m_u)m_s \Delta^2/\mu^4  $)
and the electromagnetic interaction term ($\lambda\sim\alpha^2 $).
Assuming $\alpha \sim \Delta^2/\mu^2, (m_d-m_u)m_s/\mu^2 $ (or
larger) the condition for the survival of the non-trivial minimum
is satisfied up to numerical factors. Those numerical factors
(depending, among other things, on the low energy constants
$\bar{b}$ and $\bar{c})$ determine whether the Coleman-Weinberg
mechanism occurs.
 Thus, let
us analyze them more carefully in two separate situations.

\begin{figure}[t]
\centerline{\includegraphics*[bbllx=34,bblly=587,bburx=332,bbury=792,scale=0.95,clip=true]{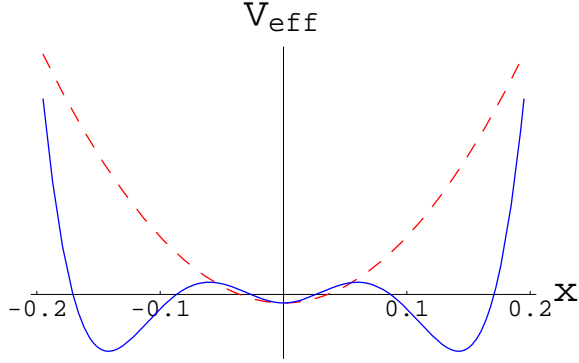}}
\caption{\label{fig:highmu}\textit Effective potential as a function of
$x=\tan(|K^+|/|K^0|$) ($\mu=10\ {\rm GeV}, \bar{b}=-1.3,
\bar{c}=0.3, $ and leading perturbative results for the remaining
parameters). The dashed line shows the effective potential without
the electromagnetic contribution. } 
\end{figure}
\noindent {\bf Asymptotic limit:}\ To analyze the simultaneous
condensation of neutral and charged kaons we use the
parameterization
\begin{eqnarray}\label{parametrization}
  K_0&=& \phi \cos x e^{i\theta_1}\nonumber\\
  K^+&=& \phi \sin x e^{i\theta_2}.
\end{eqnarray}
Due to charge and hypercharge invariance the potential does not
depend on the phases $\theta_1$ and $\theta_2$. Isospin breaking
effects create a dependence on $x$. The effective potential
(\ref{Veff}) becomes
\begin{eqnarray}\label{Veffx}
  V_{eff}&=&(m_d-m_u) \left(
  \frac{\hat{m}m_s^2f^2}{4\mu^2}-\frac{a}{8\pi}m_s\Delta^2\right)\cos 2x
  -\frac{\bar{b}}{8\pi}(\frac{2v^2}{\sqrt{\epsilon}}
+\frac{1}{v\epsilon^2})\widetilde{\alpha}
  \Delta^2 f^2 \chi\nonumber\\
  &+&(\frac{2v^4}{\sqrt{\epsilon}}
+\frac{v}{\epsilon^2})\frac{\widetilde{\alpha}^2 f^4}{16} 
  \chi^2 \ln\left( \frac{v^2 e^2 f^2}{8\Delta^2 e^{\bar{c}+\frac{3}{2}}}
  \chi\right)+ {\mathcal{O}} (\frac{(m_d-m_u)^2}{m_s^2})
\end{eqnarray}
\noindent with $\chi=2\sin^2x(\cos^2x
  +(1+\cos(\frac{\phi}{\sqrt{2}f}))\sin^2x) $ and 
 we approximate $\cos(\frac{\sqrt{2}\phi}{f})\simeq 0$.

 Let us now consider the limit $\mu\rightarrow\infty$. For
values of $\mu$ such that
\begin{equation}\label{masssmallcondition}
  \delta m \ m_s \ll \frac{\widetilde{\alpha}}{\sqrt{\epsilon}}f^2
\end{equation}
(but still satisfying the condition for $K^0$ condensation in
Eq.~(\ref{rangeDelta}), that is violated only around $10^6$ GeV)
the first term in Eq.~(\ref{Veffx}) is smaller than the second one
and can be disregarded. Numerically, condition
(\ref{masssmallcondition}) is satisfied for $\mu>3$GeV.
 Let us
momentarily put aside the second the term  in Eq.~~(\ref{Veffx})
(electromagnetic mass). Minimizing in relation to $x$ we find a
solution
\begin{equation}\label{xmin}
  x\simeq \frac{\Delta \ e^{\frac{\bar{c}}{2}+\frac{3}{4}}}{v e f}.
\end{equation}

At this value of $x$, the electromagnetic mass term can be
disregarded compared to the one we kept if
\begin{equation}\label{bsmallcondition}
  -4\bar{b}\ e^{-(c+\frac{3}{2})} \alt  1.
\end{equation}

For many, but not all, natural values of $\bar{b}$ and $\bar{c}$
this condition is satisfied and the solution in Eq.~(\ref{xmin})
can be trusted. Unfortunately, the asymptotic values of these
parameters in the limit $\mu\rightarrow\infty$ are not known (the
computation of  $\bar{c}$ involves the calculation of four loops
diagrams) and we cannot determine whether
(\ref{bsmallcondition}). In Fig.(\ref{fig:highmu}) we show, as an example, the
effective potential for a natural choice of parameter values and
very high value of the chemical potential ($\mu=10$ Gev, $m_u=4\
{\rm MeV}, m_d=7\ {\rm MeV}, m_s=150\ {\rm Mev}, \bar{b}=-1.3,
\bar{c}=0.3$). It shows the characteristic shape of  a potential
with a first order phase transition.

\begin{figure}[t]
\centerline{\includegraphics*[bbllx=26,bblly=595,bburx=300,bbury=782,scale=0.95,clip=true]{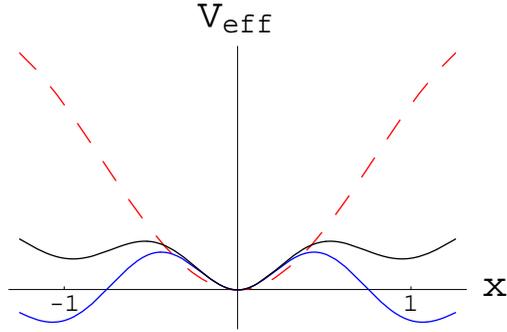}}
\caption{\label{fig:lowmu}\textit Effective potential as a function of
$x=\tan(|K^+|/|K^0|)$  ($\mu=500 \MeV,
\Delta=50 \MeV, f=3 f_{pert},  \bar{b}=-0.5,
\bar{c}=2.5$ (lower curve) and $1.5$ (upper curve)  and leading perturbative 
results for the remaining
parameters). The dashed line shows the effective potential without
the electromagnetic contribution. } 
\end{figure}

\noindent {\bf ``Realistic" densities:}\ For $\mu < 3$ Gev the
quark mass terms are no longer negligible compared to the
electromagnetic mass terms. In fact, for the densities that may be
found in neutron star cores ($\mu \simeq 500$ MeV) it is the
dominant mass term for the charged kaons and the one loop effects
are too small to overcome it for most values of the
 parameters. However, at  lower densities  the values of the low
 energy constants are not so well determined since the
 perturbative results do not apply. Some choices for the values of these
 low energy constants that do not  violate the
 expectations of dimensional analysis result in charged kaon
 condensation. As an example we show in Fig.~(\ref{fig:lowmu}) the 
effective potential for
 two choices of the parameters.

 In both of them the
  value of the decay constant $f$
 was changed from the value suggested from perturbation theory
 ($f\rightarrow 0.6 \mu = 3\ f_{pert}$) and used $\bar[b]=-0.5$.
 The two solid curves
 correspond to $\bar{c}=2.5$ and $\bar{c}=1.5$. This change in
 $\bar{c}$ is
 enough to transform the global minimum into a local minimum.
 This set of parameters were carefully chosen For most of the
 parameter space the quark mass term overwhelms the others and
 there is no charged kaon condensation.

A better idea of the likelihood of charged kaon condensation at
these densities can be perhaps obtained through the use of QCD
models to estimate the unknown low energy constants in the density range 
inaccessible to perturbation theory.

We have considered the possibility of charged kaon condensation
 and (electromagnetic) superconductivity at high dense quark
 matter. At asymptotically high densities, where perturbative QCD
 applies and the question can be decided on first principles, a
 complicated computation of some low energy constants are
 necessary to settle the issue. We find however that for most natural 
values of these
 constants charged kaon condensation indeed occurs. At lower
 densities the situation is the opposite. For most reasonable values of the
low energy constants the quark mass effects overwhelm the
electromagnetic effects and there is no $K^+$ condensation.

\begin{acknowledgments}
I would like to acknowledge useful discussions with D. Kaplan and
 T. Sch\"afer.
This work was supported by the Director, Office of Energy Research, 
Office of High Energy and Nuclear Physics, and by the Office of 
Basic Energy Sciences, Division of Nuclear Sciences, 
of the U.S. Department of Energy under Contract No.
DE-AC03-76SF00098. 
\end{acknowledgments}


\end{document}